\documentclass{aa}  

\usepackage{graphicx}
\usepackage{txfonts}

\begin{document}

\titlerunning{Gradual solar energetic particle event with enhanced \element[ ][3]{He} abundance}

   \title{The first gradual solar energetic particle event with enhanced \element[ ][3]{He} abundance on Solar Orbiter}


 \author{R. Bu\v{c}\'ik\inst{1}
          \and
          G. M. Mason\inst{2}
          \and
          R. G\'omez-Herrero\inst{3}
          \and   
          V. Krupar\inst{4,5}
          \and
          D. Lario\inst{4}
          \and
          M. J. Starkey\inst{1}
          \and
          G. C. Ho\inst{2}
           \and
         J. Rodr\'iguez-Pacheco\inst{3}
         \and
          R. F. Wimmer-Schweingruber\inst{6}
          \and
           F. Espinosa Lara\inst{3}
           \and
           T. Tadesse\inst{7}
           \and
          L. Balmaceda\inst{4,8}
          \and
          C. M. S. Cohen\inst{9}
            \and
          M. A. Dayeh\inst{1}
            \and
          M. I. Desai\inst{1}
            \and
           P. K\"uhl\inst{6}
             \and
          N. V. Nitta\inst{10}
          \and
        M. E. Wiedenbeck\inst{11}
        \and
        Z. G. Xu\inst{6}
 }

   \institute{Southwest Research Institute, San Antonio, TX 78238, USA\\
              \email{radoslav.bucik@swri.org}
         \and
            Applied Physics Laboratory, Johns Hopkins University, Laurel, MD 20723, USA
            \and
            Universidad de Alcal\'a, Space Research Group, 28805 Alcal\'a de Henares, Spain
            \and
            Heliophysics Science Division, NASA Goddard Space Flight Center, Greenbelt, MD, USA 
            \and
            Goddard Planetary Heliophysics Institute, University of Maryland, Baltimore, MD 21250, USA 
           \and
            Institut f\"{u}r Experimentelle und Angewandte Physik, Christian-Albrechts-Universit\"{a}t zu Kiel, Kiel, Germany 
            \and
            NASA Johnson Space Center, Houston, TX, USA
            \and
            George Mason University, Fairfax, VA, USA 
             \and
             California Institute of Technology, Pasadena, CA 91125, USA 
             \and
           Lockheed Martin Advanced Technology Center, Palo Alto, CA 94304, USA
            \and
           Jet Propulsion Laboratory, California Institute of Technology, Pasadena, CA 91109, USA
             }

   \date{Received ; accepted }

 
  \abstract{

The origin of $^3$He abundance enhancements in coronal mass ejection (CME)-driven shock gradual solar energetic particle (SEP) events remains largely unexplained. Two mechanisms have been suggested -- the re-acceleration of remnant flare material in interplanetary space and concomitant activity in the corona. We explore the first gradual SEP event with enhanced $^3$He abundance observed by Solar Orbiter. The event started on 2020 November 24 and was associated with a relatively fast halo CME. During the event, the spacecraft was at 0.9\,au from the Sun. The event averaged $^3$He/$^4$He abundance ratio is 24 times higher than the coronal or solar wind value, and the $^3$He intensity had timing similar to other species. We inspected available imaging, radio observations, and spacecraft magnetic connection to the CME source. It appears the most probable cause of the enhanced $^3$He abundance are residual $^3$He ions remaining from a preceding long period of $^3$He-rich SEPs on 2020 November 17--23.
}

   \maketitle
%

\section{Introduction}

Solar energetic particles (SEPs) are produced by mechanisms related to magnetic reconnection in flares or jets in Impulsive SEP (ISEP) events and by coronal mass ejection (CME)-driven shocks traveling through the corona and interplanetary space in Gradual SEP (GSEP) events \citep[e.g.,][]{2021LNP...978.....R}. One of the most strikingly distinct characteristics between these groups is their different elemental composition \citep[e.g.,][]{2021SSRv..217...72R}. ISEP (or $^3$He-rich) events have enhanced $^3$He and heavy-ion abundances compared to corona or solar wind values \citep[e.g.,][]{2007SSRv..130..231M}, while GSEP events have abundances similar to coronal values \citep[e.g.,][]{2016LRSP...13....3D}. High-resolution mass spectrometers onboard the Advanced Composition Explorer (ACE) have often revealed small enhancements of $^3$He and Fe abundances in GSEP events \citep[e.g.,][]{1999GeoRL..26..141M,1999ApJ...525L.133M,1999GeoRL..26.2697C,2000AIPC..528..107W,2016ApJ...816...68D}, thus blurring the distinction between jet- and CME-related SEPs. While in GSEP events the Fe/O enhancement can result from magnetic rigidity-dependent transport effects due to the smaller charge-to-mass ratio of Fe compared to O, such an effect is not expected for $^3$He/$^4$He enhancement \citep[e.g.,][]{1999ApJ...525L.133M}. Thus, a $^3$He enhancement unambiguously indicates the presence of flare material in GSEP events \citep[e.g.,][]{2016ApJ...816...68D}.     

\citet{1999GeoRL..26..141M} reported two GSEP events on 1997 November 4 and November 6, that had a 0.5--2.0\,MeV\,nucleon$^{-1}$  $^3$He/$^4$He ratio factor of 4 above the average solar wind value of (4.8$\pm$0.5)$\times$10$^{-4}$ \citep{1980JGR....85.6021O}. \citet{1999ApJ...525L.133M} examined 12 GSEP events and found a finite 0.5--2.0\,MeV\,nucleon$^{-1}$ $^3$He peak in eight events with a $^3$He/$^4$He ratio between 4.7 and 135 times the average slow solar wind $^3$He/$^4$He of (4.08$\pm$0.25)$\times$10$^{-4}$ \citep{1998SSRv...84..275G}. The 0.4--0.6\,MeV\,nucleon$^{-1}$ Fe/O$\sim$0.79 was also enhanced compared to the average GSEP value. The five events in \citet{1999ApJ...525L.133M} showed a similar $^3$He and $^4$He time-intensity profiles, indicating a common acceleration and transport origin. These authors suggested that residual $^3$He ions from ISEP events, which was found to be commonly present in the interplanetary medium at 1\,au, form a source material for $^3$He seen in GSEP events. Similarly, \citet{2005ApJ...625..474T} proposed that enhanced Fe/O in GSEP events may be produced by the acceleration of remnant suprathermal Fe-rich ion population from previous small flare events. \citet{2016ApJ...816...68D} identified 27 out of 46 GSEP events with a finite $^3$He mass peak and a 0.5--2.0\,MeV\,nucleon$^{-1}$ $^3$He/$^4$He ratio between $\sim$1.5 and $\sim$194 times the average slow solar wind value \citep{1998SSRv...84..275G}. A flare component simultaneously produced in the parent active regions (ARs) \citep{2000AIPC..528..111V,2003GeoRL..30.8017C,2006JGRA..111.6S90C,2013ApJ...776...92K} or in other sites in the corona \citep{2013ApJ...776...92K} has been proposed as causes of the Fe/O and $^3$He/$^4$He enhancement in GSEP events. The longitude of the observer relative to source AR \citep{2003GeoRL..30.8017C} and the nature of the photospheric field where the observer is magnetically connected to \citep{2013ApJ...776...92K} have also been suggested to affect the Fe/O ratio.  

The unresolved question of whether residual flare material in interplanetary space or concomitant coronal activity dominates $^3$He abundance enhancements in GSEP events could be thoroughly explored with new high-resolution in-situ and remote measurements from Solar Orbiter \citep{2020A&A...642A...1M} at close distances from the Sun. In this study, we investigate the first GSEP event with enhanced $^3$He abundance measured by Solar Orbiter and analyze the origin of the observed $^3$He enhancement. The event started on 2020 November 24. It is also the first significant SEP event observed by this mission in the sense of high proton intensities. At suprathermal energies ($<$1\,MeV\,nucleon$^{-1}$), this event shows larger proton intensities than the following SEP event on 2020 November 29 \citep{2021A&A...656A..20K}, which was observed at widely ($\sim$230$^{\circ}$) separated spacecraft and associated with the same AR as the 2020 November 24 event.


\section{Instrumentation}

The 2020 November 24 event abundance measurements were obtained by the Suprathermal Ion Spectrograph (SIS) of the Energetic Particle Detector (EPD) suite \citep{2020A&A...642A...7R} aboard Solar Orbiter. SIS is a time-of-flight mass spectrometer that measures elemental composition from H through ultra-heavy nuclei in the kinetic energy range of $\sim$0.1--10\,MeV\,nucleon$^{-1}$. SIS has two telescopes, one (SIS-A) pointing at 30$^{\circ}$ (sunward) and the other (SIS-B) at 160$^{\circ}$ (anti-sunward) to the west of the spacecraft-Sun line. The energetic proton measurements were obtained from the High-Energy Telescope (HET) of the EPD suite. HET has four viewing directions, with one unit pointing sun- and anti-sunward along the average Parker spiral and the other unit pointing out of the ecliptic. We also used electron measurements from EPD made by the SupraThermal Electron Proton (STEP), Electron Proton Telescope (EPT), and HET instruments. To determine the interplanetary magnetic field (IMF) polarity, we utilized magnetic field data from the magnetometer (MAG) on board Solar Orbiter \citep{2020A&A...642A...9H}.

The coronal activity associated with the event was examined using EUV and white-light images from the EUVI instrument and COR2 coronagraph on STEREO-A \citep{2008SSRv..136...67H}, respectively. The Solar Orbiter EUV imaging observations from the Extreme-Ultraviolet Sensor Imager \citep[EUI;][]{2020A&A...642A...8R} were not available for the 2020 November 24 event. Further, we inspected radio data from the STEREO-A/WAVES \citep{2008SSRv..136..487B} and the Solar Orbiter Radio and Plasma Waves \citep[RPW;][]{2020A&A...642A..12M} instruments.

\section{Observations}

  \begin{figure}
  \centering
      \includegraphics[width=0.48\textwidth]{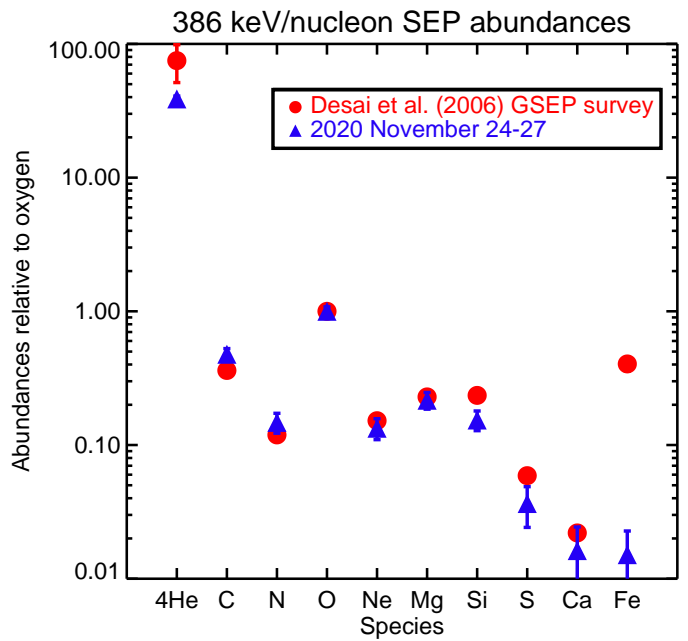}
   \caption{Abundances relative to O for the 2020 November 24 event (blue triangles). Reference ion abundances (red circles) are from the GSEP event survey by \citet{2006ApJ...649..470D}.}
              \label{f0}%
    \end{figure}

The selection of the examined event as a GSEP event is primarily based on abundance ratios that are comparable with the reference population. Furthermore, the event is associated with a large and fast CME, which is typically observed in GSEP events. Figure~\ref{f0} displays an abundance plot for the event and reference population obtained in the GSEP survey by \citet{2006ApJ...649..470D}.

\subsection{Solar energetic ions}

   \begin{figure*}
   \centering
     \includegraphics[width=1.\textwidth]{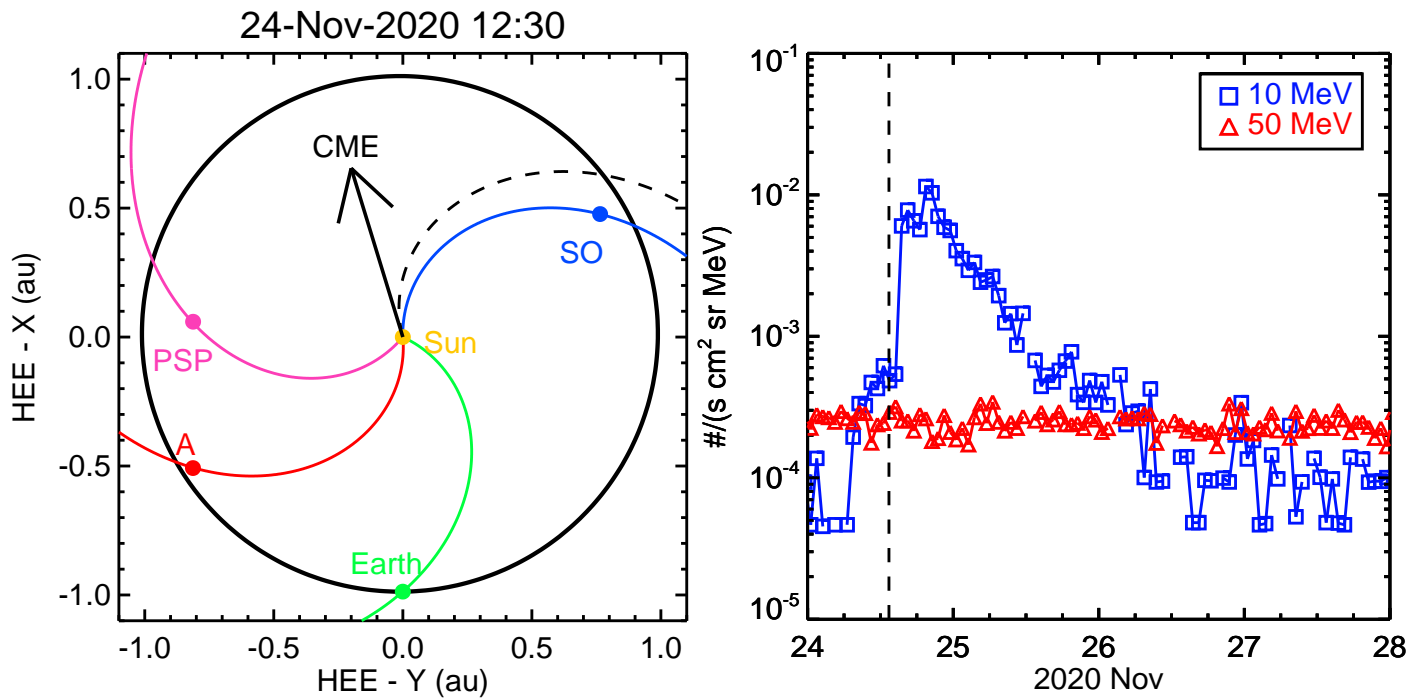}
   \caption{Constellation of selected spacecraft and high-energy proton observation of GSEP event. (Left) Blue, pink, red, and green circles mark the Heliocentric Earth Ecliptic (HEE) location of Solar Orbiter (SO), Parker Solar Probe (PSP), STEREO-A (A), and Earth, respectively on 2020 November 24 at 12:30\,UT. The arrow indicates the estimated longitude of the CME source. Nominal Parker spiral IMF lines assuming a 350\,km\,s$^{-1}$ solar wind speed are shown. (Right) Solar Orbiter EPD/HET 1-hr omnidirectional proton intensity, obtained by averaging intensities of the four telescopes. The curve with blue squares, marked by 10\,MeV, is a combination of 11 energy bins between 7.0 and 11.3\,MeV; the curve with red triangles, marked by 50 MeV, is a combination of 13 energy bins between 27.2\,MeV and 75.1\,MeV. The vertical dashed line indicates the first appearance of the CME in the STEREO-A/COR2 coronagraph.}
              \label{f1}%
    \end{figure*}

Figure~\ref{f1}~(Left) shows the locations of selected spacecraft in the plane of the ecliptic as seen from the north on 2020 November 24 at 12:30\,UT in the Heliocentric Earth Ecliptic (HEE) coordinates. The dashed Parker spiral connects to the CME source longitude (E105 with respect to STEREO-A) as inferred from EUV STEREO-A observations (see Section~\ref{Ss} below). Solar Orbiter is well connected to the CME source site along a nominal Parker spiral IMF line. The angular separation between the Solar Orbiter magnetic foot-point longitude on the Sun and the longitude of the CME source is $\sim$15$^{\circ}$ for a nominal Parker spiral for a 350\,km\,s$^{-1}$ solar wind speed. We note the solar wind measurements from the Solar Orbiter Solar Wind Analyser \citep[SWA;][]{2020A&A...642A..16O} are not available for the period of interest. The event was also observed by Parker Solar Probe (PSP) but neither by STEREO-A nor by spacecraft located near Earth (e.g., ACE). The longitudinal separation between Solar Orbiter and PSP of 144$^{\circ}$ suggests widespread SEPs in the event. Solar Orbiter was at 0.9\,au and PSP at 0.8\,au from the Sun. The EPD/HET Solar Orbiter 1-hr omnidirectional proton intensities at 10 and 50\,MeV for the examined event are displayed in Fig.~\ref{f1}~(Right). There is no measured intensity enhancement at 50\,MeV. The rapid onset of the event, seen just after the vertical dashed line marking the first appearance of the CME in the coronagraph, is preceded by a small proton enhancement at $\sim$07\,UT. There has been reports of EPD electron enhancements with onset at 06:51\,UT at 53--85\,keV \citep{2021A&A...656A..22W}, which is likely related to this small proton event.

   \begin{figure*}
   \centering
     \includegraphics[width=.91\textwidth]{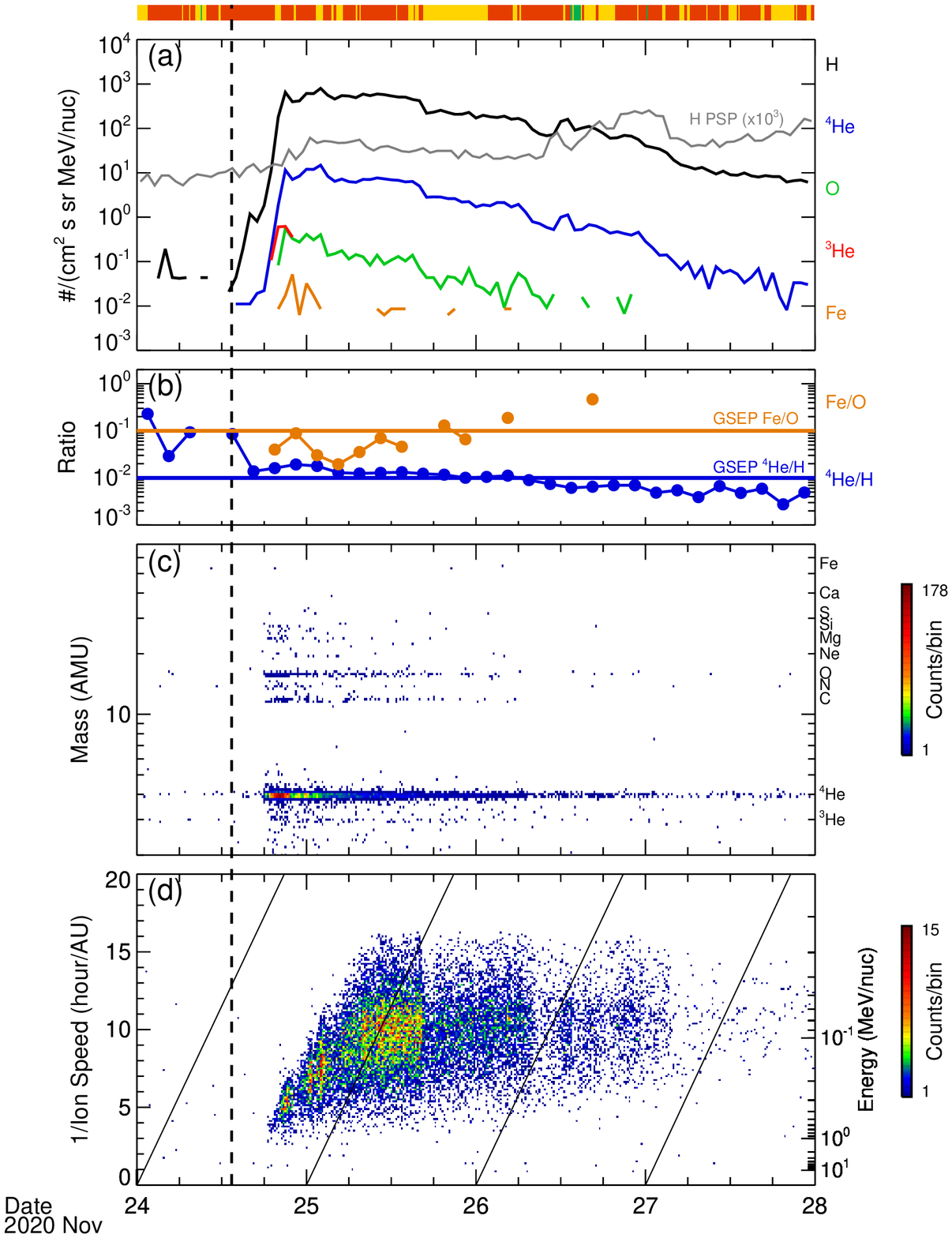}
   \caption{GSEP event with time profiles of various physical quantities. (a) The Solar Orbiter SIS 1-hr H, $^3$He, $^4$He, O, Fe intensity at 0.23--0.32\,MeV\,nucleon$^{-1}$. H PSP curve is a combination of 1-hr H count rates (s$^{-1}$) from 14 energy bins between 0.023 MeV and 0.187 MeV from PSP IS$\sun$IS EPI-Lo. (b) The Solar Orbiter SIS 3-hr $^4$He/H and Fe/O ratios at 0.23--0.32\,MeV\,nucleon$^{-1}$. The horizontal lines mark $^4$He/H and Fe/O in GSEP events \citep{1995RvGeo..33S.585R}. (c) The Solar Orbiter SIS mass vs. time at 0.4--10\,MeV\,nucleon$^{-1}$. (d) The Solar Orbiter SIS 1/(ion speed) vs. arrival times of 10--70\,AMU ions. The measurements are from both SIS telescopes and averaged together. The vertical dashed line indicates the first appearance of the CME in the STEREO-A/COR2 coronagraph. Sloped lines mark arrival times for particles traveling along the nominal Parker field line without scattering. The horizontal color bar on the top shows IMF polarity at Solar Orbiter. Red is negative, green is positive, and yellow is an ambiguous polarity. }
              \label{f2}%
    \end{figure*}

\begin{figure*}
   \centering
     \includegraphics[width=0.44\textwidth,angle=90]{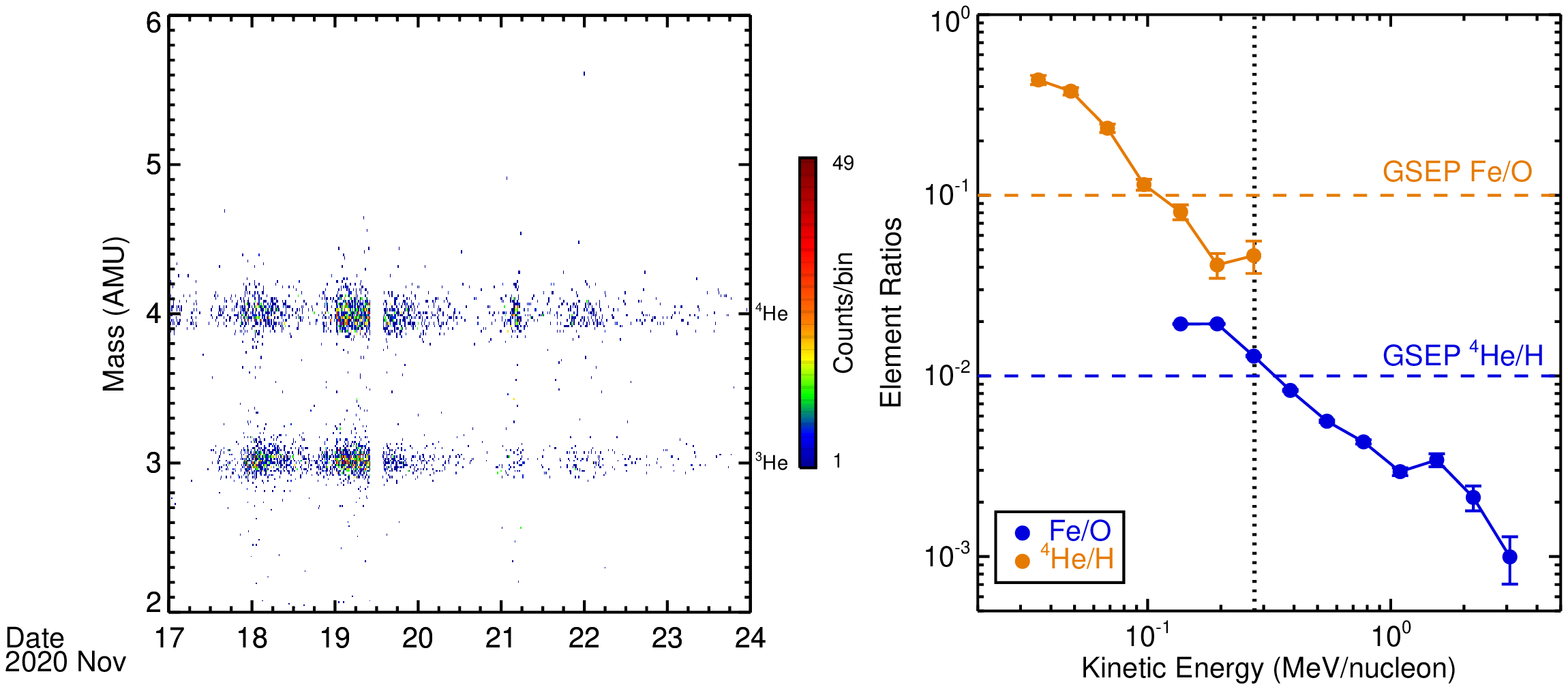}
   \caption{Long period of \element[ ][3]{He}-rich SEPs and energy dependence of selected abundance ratios. (Left) Solar Orbiter SIS He mass vs. time at 0.4--10\,MeV\,nucleon$^{-1}$. The sharp cutoff of particles around mid-day on November 19 was due to instrument maintenance. (Right) The event-integrated element ratios Fe/O and \element[ ][4]{He}/H over the period of 2020 November 24 12:00\,UT--2020 November 27 00:00\,UT vs. energy. The vertical dotted line marks the energy corresponding to ratios in Fig.~\ref{f2}b. The measurements are from both SIS telescopes averaged together.}
              \label{f3}%
    \end{figure*}

Figure~\ref{f2} shows Solar Orbiter SIS measurements of the event on 2020 November 24--27. Figure~\ref{f2}a displays 1-hr H, $^3$He, $^4$He, O, Fe intensity at 0.23--0.32\,MeV\,nucleon$^{-1}$. The 1-hr $^3$He intensity is plotted only if the 0.23--0.32\,MeV\,nucleon$^{-1}$ $^3$He/$^4$He ratio is above 5\%, and there are more than eight $^3$He counts throughout the 1-hr interval. The small H intensity enhancement, preceding the main event, is also seen at 0.23--0.32\,MeV\,nucleon$^{-1}$ around midday of November 24 which agrees with the $\sim$10 MeV onset at $\sim$07\,UT. The heavier ions had intensities too low to be detected along with H. The 2020 November 24 event was only marginally (compared to Solar Orbiter) observed on PSP with an approximate order of magnitude increase above the pre-event background. The event on PSP was clearly seen at energies below $\sim$0.2 MeV with a velocity dispersive profile. Overplotted in Fig.~\ref{f2}a with a gray line are 1-hr H count rates summed over 14 energy bins between 0.023\,MeV and 0.187\,MeV from PSP IS$\sun$IS EPI-Lo \citep{2017JGRA..122.1513H}. The elements with Z$>$ 2 were not measured on PSP in this event. The Solar Orbiter SIS $^4$He/H and Fe/O ratios at 0.23--0.32\,MeV\,nucleon$^{-1}$ shown in Fig.~\ref{f2}b are close to GSEP events values. The $^4$He/H and Fe/O in ISEP events are about ten times higher \citep[e.g.,][]{1995RvGeo..33S.585R}. The Solar Orbiter SIS mass spectrogram (Fig.~\ref{f2}c) shows a clear $^3$He presence and almost lack of Fe during the event at energy 0.4--10\,MeV\,nucleon$^{-1}$. A few $^3$He ions detected on November 24, just before the event may be remnants from a long period of $^3$He-rich SEPs measured with SIS on Solar Orbiter on November 17--23 \citep[see the left panel of Fig.~\ref{f3} and also][]{2021A&A...656L..11B}. The 1/ion-speed vs. time plot (Fig.~\ref{f2}d) shows a dispersive triangular pattern where a rough extrapolation to zero of the inverted speed suggests an ion solar release time around 13:00\,UT. Previous studies reported uncertainty of $\pm$45 minutes in estimates of release times from low energy $<$1\,MeV\,nucleon$^{-1}$ measurements \citep[e.g.,][]{2000ApJ...545L.157M}. The horizontal color bar on top of Fig.~\ref{f2} shows IMF polarity determined from 10-minute Solar Orbiter MAG data. Red denotes IMF azimuthal angles 90$^{\circ}$--180$^{\circ}$ which typically correspond to toward or negative polarity. Green denotes angles 270$^{\circ}$--360$^{\circ}$ (away or positive polarity). Yellow marks remaining angles that have an ambiguous polarity.

Figure~\ref{f3}~(Left) displays the He mass spectrogram for the period 2020 November 17--23 \citep{2021A&A...656L..11B}. Figure~\ref{f3}~(Right) shows the event-integrated $^4$He/H (blue) and Fe/O (orange) as a function of energy. Note that the Fe/O spectrum only extends to $\sim$0.3\,MeV\,nucleon$^{-1}$ due to the lack of Fe counts above 0.4\,MeV\,nucleon$^{-1}$ (Fig.~\ref{f2}b).  The $^4$He/H is close to or lower than the GSEP value of $\approx$0.01 \citep[e.g.,][]{1995RvGeo..33S.585R,2021ApJ...908..214K}. The Fe/O is close to or higher than the GSEP value of $\approx$0.1 but does not reach the ISEP value of $\approx$1 \citep[e.g.,][]{1995RvGeo..33S.585R}. A decrease of $^4$He/H and Fe/O with increasing kinetic energy has been previously reported in GSEP events \citep[e.g.,][]{2016LRSP...13....3D}.

  \begin{figure*}
  \centering
      \includegraphics[width=0.44\textwidth,angle=90]{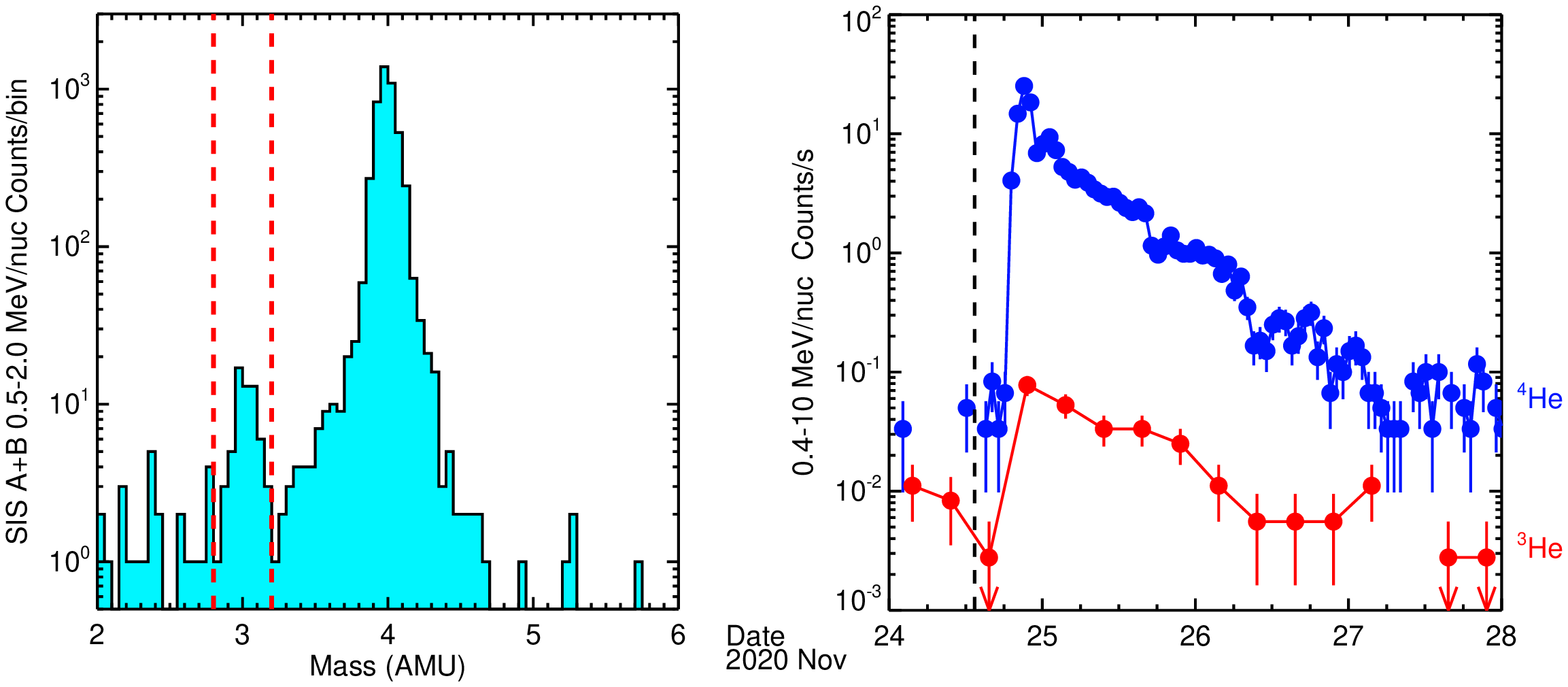}
   \caption{$^3$He mass peak and comparison of $^3$He and $^4$He counts\,s$^{-1}$ time profiles. (Left) He mass histogram from both SIS telescopes A and B for 2020 November 24 12:00\,UT--2020 November 27 00:00\,UT and energy range of 0.5--2.0\,MeV\,nucleon$^{-1}$. The red vertical dashed lines mark the mass range for $^3$He count rates shown in the right panel. (Right) The SIS 1-hr $^4$He and 6-hr $^3$He count rates at 0.4--10\,MeV\,nucleon$^{-1}$. The black vertical dashed line indicates the first appearance of the CME in the coronagraph.}
              \label{f4}%
    \end{figure*}
    
The He mass histogram in Fig.~\ref{f4}~(Left) shows a clear $^3$He peak well separated from the $^4$He peak. The event-integrated 0.5--2.0\,MeV\,nucleon$^{-1}$ $^3$He/$^4$He ratio of 0.0094$\pm$0.0014 is $\sim$23.5 times higher than the average $^3$He/$^4$He of (4.08$\pm$0.25)$\times$10$^{-4}$ in the slow solar wind \citep{1998SSRv...84..275G}. Figure~\ref{f4}~(Right) shows time profiles of 0.4--10\,MeV\,nucleon$^{-1}$ 1-hr $^4$He and 6-hr $^3$He count rates. We use 2.8--3.2\,AMU and 3.3--6\,AMU mass ranges for $^3$He and $^4$He count rates, respectively. $^4$He data points with one count are not plotted. $^3$He data points with one count are shown and have error bars marked by an arrow. The $^3$He count rate time profile is similar to the $^4$He time profile except for a few $^3$He points with large errors near the end of the event. It has been argued that such similarity suggests the same acceleration and propagation histories \citep[e.g.,][]{1999ApJ...525L.133M,2000AIPC..528..107W}. Figure~\ref{f4}~(Right) also shows that $^3$He count rates increase significantly compared to the pre-event $^3$He background. 

 \begin{figure*}
  \centering
     \includegraphics[width=1.\textwidth]{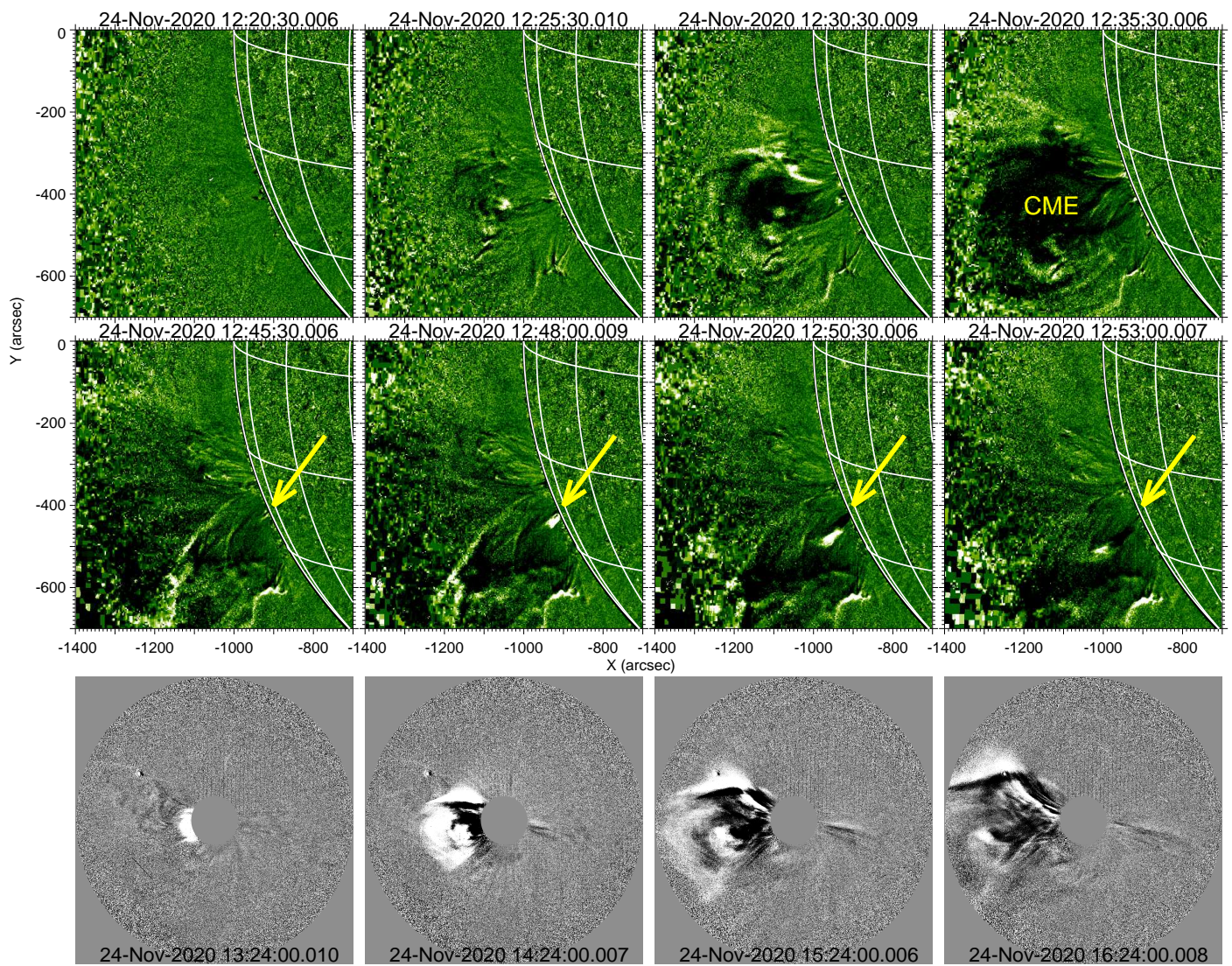}
   \caption{CME and jet expansion. (Top, Mid) STEREO-A EUVI 2.5-minute running difference EUV 195\,{\AA} images. The arrows mark the post-CME jet-like ejection. The heliographic longitude-latitude grid has 15\degr spacing. (Bottom) STEREO-A COR2 30-minute running difference white-light images.}
              \label{f5}%
    \end{figure*}

\begin{figure*}
  \centering
     \includegraphics[width=0.252\textwidth, angle=90]{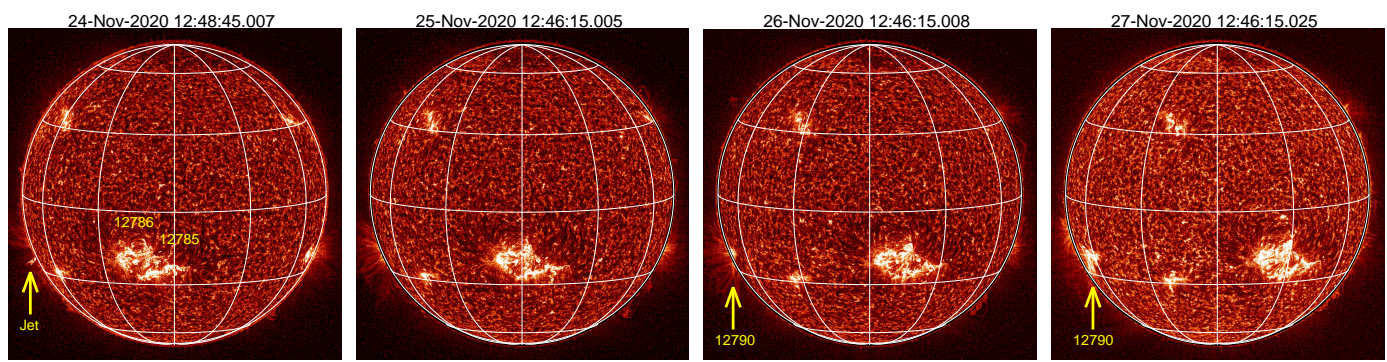}
   \caption{CME source AR and jet in full-disk solar images. STEREO-A EUVI direct EUV 304\,{\AA} images. The heliographic longitude-latitude grid has 30$\degr$ spacing.}
              \label{f6}%
    \end{figure*}

Figure~\ref{f6} shows full-disk STEREO-A EUV daily images between 2020 November 24 and November 27. The above-mentioned jet is marked in the leftmost panel. The large bright area that passed through the central meridian represents AR 12786 and 12785, the sources of a long period of  $^3$He-rich SEPs \citep{2021A&A...656L..11B}. The AR 12790, where the CME (and probably the jet) originated, is marked in the two rightmost panels.

\subsection{Solar source} \label{Ss}%

The examined SEP event is related to the partial halo CME with a reported speed of 892\,km\,s$^{-1}$ and width of 106$\degr$ that appeared in the COR2 coronagraph on board STEREO-A on 2020 November 24 at 13:24\,UT, consistent with the estimated ion injection time at $\sim$13\,UT. These values are from the STEREO-A automatically generated list of CMEs\footnote{https://secchi.nrl.navy.mil/cactus/} \citep[][]{2009ApJ...691.1222R} that utilizes images from COR2 coronagraph (field of view 2.5--15\,R$_{\sun}$). The SOHO LASCO manually-identified CMEs catalog\footnote{https://cdaw.gsfc.nasa.gov/CME$\_$list/} \citep[][]{2004JGRA..109.7105Y} reports a halo CME with a speed of 668\,km\,s$^{-1}$ that appeared in C2 coronagraph (1.5--6\,R$_{\sun}$) on 2020 November 24 at 13:25\,UT. Figure~\ref{f5}~(Top row) shows the CME expulsion between 12:20 and 12:35\,UT on November 24 in STEREO-A EUV images from EUVI that has a field of view extending to 1.7\,R$_{\sun}$. The CME eruption started around 12:25\,UT from the area near $\sim$S20 at the east limb as viewed from STEREO-A. In the same area, STEREO-A EUVI began to observe a small CME post-eruptive arcade (not shown) at $\sim$14:00\,UT on November 24, indicating that the CME source was close to the east limb. On November 26, STEREO-A spotted the responsible AR (number 12790), with an estimated location of S25E105 (from STEREO-A viewpoint) at the time of the CME eruption. There was a jet-like eruption, $\sim$20\,min after CME, apparently from the same area and projected direction as the CME (Fig.~\ref{f5}, Mid row arrow). Figure~\ref{f5}~(Bottom row) shows the progression of the CME in the COR2 coronagraph.

\begin{figure}
  \centering
      \includegraphics[width=0.48\textwidth]{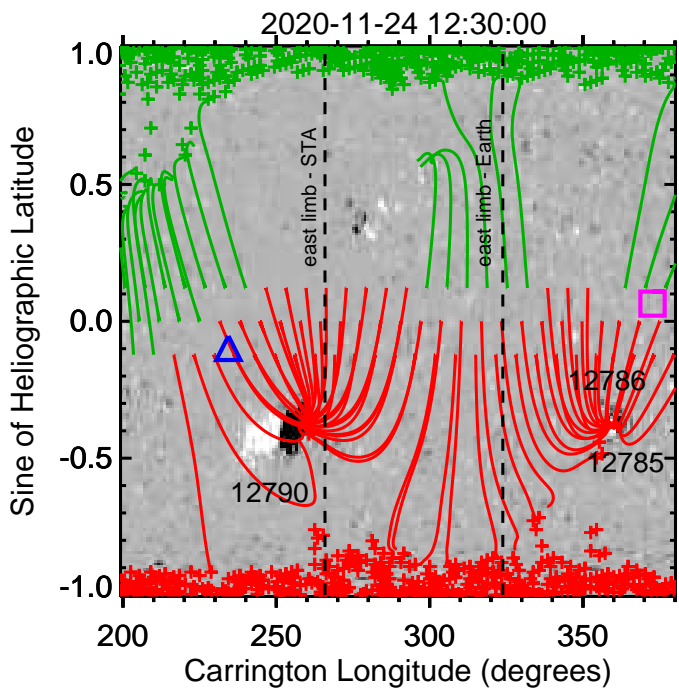}
   \caption{Magnetic field map with coronal field model. Photospheric magnetic field, scaled to $\pm$30\,Mx\,cm$^{-2}$ (grayscale), and PFSS model coronal field (red-negative and green-positive polarity). Shown are field lines that intersect the source surface at latitudes 0\degr and  $\pm$7\degr. The blue triangle marks Solar Orbiter, and the pink square marks PSP magnetic foot-points at the source surface. Two dashed vertical lines mark Carrington longitudes of the east solar limb as viewed from STEREO-A (STA) and the Earth.}
              \label{f7}%
    \end{figure}

Figure~\ref{f7} displays a photospheric magnetic field map with a Potential Field Source Surface \citep[PFSS;][]{1969SoPh....6..442S} model coronal field around the onset of the CME. The PFSS model presents a simple approximation to the real global coronal field, which neglects the effects of the forces and electric currents. The PFSS model was computed using the SolarSoft PFSS package that uses the SDO HMI line-of-sight magnetograms, which are assimilated into the flux-dispersal model to provide the magnetic field on the full solar sphere \citep{2003SoPh..212..165S}. The source surface that separates open and closed fields was set to a commonly used 2.5\,R$_{\sun}$ from the Sun’s center. Figure~\ref{f7} shows that the Solar Orbiter magnetic foot-point longitude (blue triangle), for a nominal solar wind speed of 350\,km\,s$^{-1}$, was close to the longitude of the CME source AR 12790. Furthermore, Solar Orbiter could be connected to AR 12790 via open negative polarity coronal field lines emanating from the AR. We note the in-situ IMF polarity on Solar Orbiter (see Fig.~\ref{f2}) is consistent with the model prediction. The PSP foot-point (pink square) separation of $\sim$120$\degr$ from the CME source AR could be the reason that only a small intensity increase was observed on the spacecraft. 

\begin{figure*}
  \centering
      \includegraphics[width=0.78\textwidth, angle=-90]{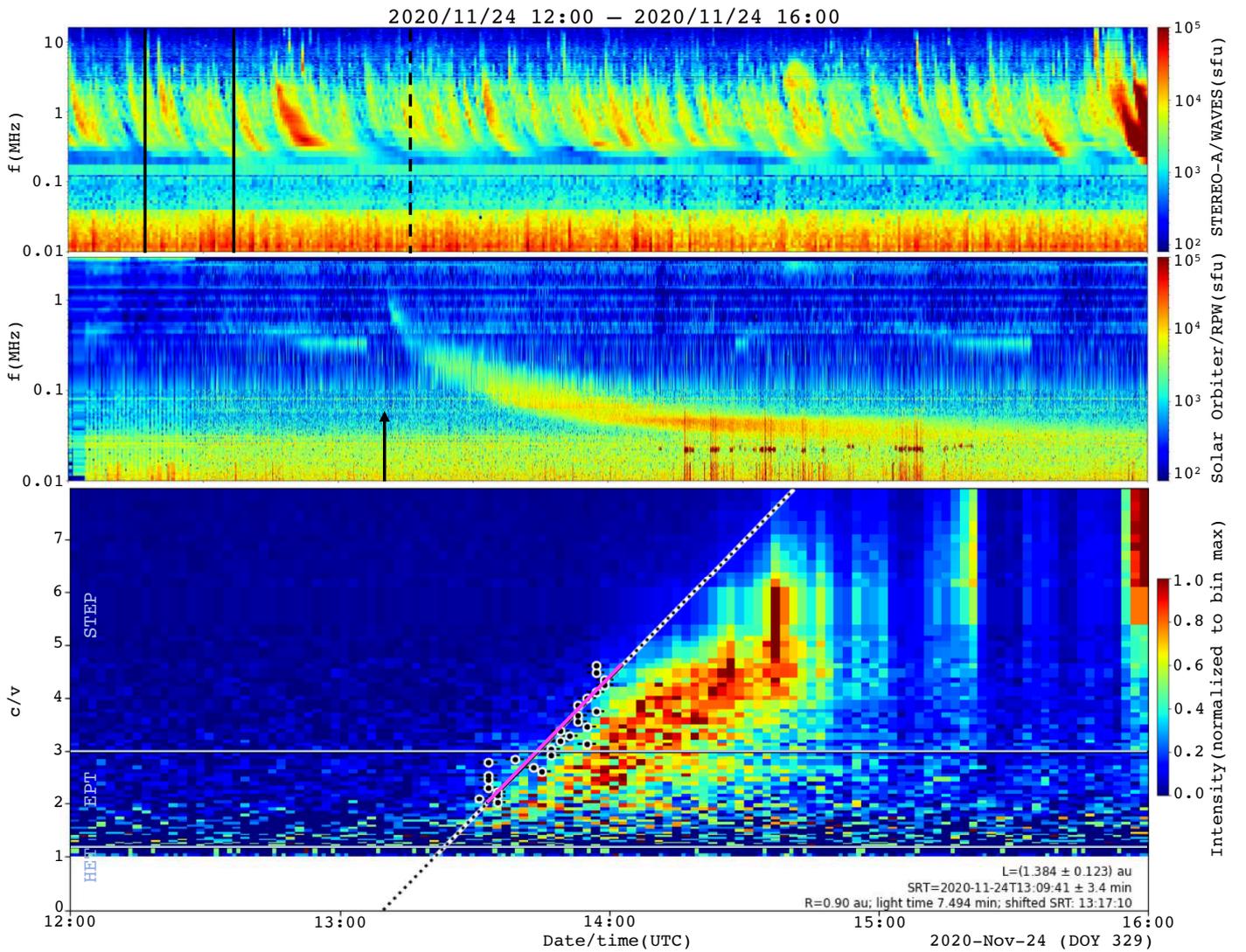}
   \caption{Radio and electron spectrograms. (Top) STEREO-A WAVES and (Mid) Solar Orbiter RPW radio spectrograms with times shifted to Sun. Two vertical solid lines mark the start of the CME and jet-like eruptions in EUVI, respectively. The vertical dashed line marks the first appearance of the CME in COR2. The arrow marks the onset of type III radio bursts at 13:18\,UT observed by Solar Orbiter. We note that all vertical lines in the top panel and the arrow in the middle panel mark times shifted to the Sun. (Bottom) Electron $c/v$ vs. time plot, where $c$ is the speed of light and $v$ is electron velocity. The slanted line is a linear fit to the electron arrival times; the solid part of the line marks the energy range used for the fit. Gray horizontal lines separate STEP, EPT, and HET energy ranges.}
              \label{f8}%
    \end{figure*}

Figure~\ref{f8} (Top and Mid) displays STEREO-A/WAVES and Solar Orbiter/RPW radio spectrograms with times shifted to Sun covering a period of the CME and the jet occurrence. The two spectrograms show a markedly different appearance. STEREO-A saw storm type III bursts at a frequency range from $\sim$5\,MHz to $\sim$300\,kHz. The jet start time, shifted to Sun, is marked by the second vertical solid line in the top panel. There is very likely a coincidental observation of a storm type III at 12:45\,UT with the jet at STEREO-A, otherwise, we would see this type III burst also on Solar Orbiter. Thus, no clear type III burst association with the solar jet can be found. In the period covered by Fig.~\ref{f8}, Solar Orbiter observed a single type III radio burst starting at 13:18\,UT (or at 13:11\,UT shifted to Sun) and drifting from $\sim$2\,MHz \citep[corresponding to $\sim$6\,R$_{\sun}$;][]{1999A&A...348..614M} to the local plasma frequency ($\sim$30\,kHz). The start time of the type III burst, marked by an arrow in the middle panel, closely precedes the first appearance of the CME in the COR2, marked by the vertical dashed line in the top panel. Figure~\ref{f8}~(Bottom) displays Solar Orbiter electron $c/v$ vs. time spectrogram, formed using, from top to bottom, STEP, EPT and HET data, showing dispersive energetic electron event. Although the observations are limited by the relatively low statistics, the inferred electron solar release time (SRT) at 13:09:41\,UT well agrees with the type III burst onset shifted to Sun at 13:11\,UT.

Type III storms were observed by STEREO-A throughout 2020 November 23--29. They probably originated in closely spaced ARs 12785 and 12786 (see Fig.~\ref{f6}) that showed a persistent brightening. As these regions rotated out of the STEREO-A view, the storm activity decreased.

\section{Discussion}

In this paper, we examine the first intense GSEP event with enhanced $^3$He abundance measured on Solar Orbiter. We find $^3$He abundance significantly (by a factor of $\sim$24) enhanced above the coronal or solar wind abundance. The fundamental question concerning the origin of the $^3$He in the examined event is whether this $^3$He enhancement is due to remnant material left in the heliosphere or simultaneous activity in the corona (e.g., parent AR).

A timeline of solar events is summarized in Table~\ref{t1}. Column 1 shows the type of activity. The instruments where the activity was observed are given in parentheses. Columns 2 and 3 give the observed approximate times and times shifted to the Sun, respectively.

\begin{table*}
\caption{\label{t1}A timeline of solar events.}
\centering
\begin{tabular}{c c c}
\hline\hline
 Activity&Observed time (UT)&Time at Sun (UT)\\
\hline
CME start (EUVI)&2020-Nov-24 12:25&2020-Nov-24 12:17\\
Jet start (EUVI)&2020-Nov-24 12:45&2020-Nov-24 12:37\\
Ion injection (EPD)&...&2020-Nov-24 13:00\\
Electron injection (EPD)&...&2020-Nov-24 13:09\\
Type III burst (RPW)&2020-Nov-24 13:18&2020-Nov-24 13:11\\
CME appearance (COR2)&2020-Nov-24 13:24&2020-Nov-24 13:16\\
\hline
\end{tabular}
\end{table*}

A jet-like expulsion (see Fig.~\ref{f5}) was observed to closely follow (after $\sim$20 minutes) the CME liftoff. If $^3$He-rich SEPs were produced in association with this jet, the ions with kinetic energy, e.g., 0.23--0.32\,MeV\,nucleon$^{-1}$ would arrive at Solar Orbiter around $\sim$19--20\,UT (assuming a nominal Parker spiral), which is consistent with the timing of measurements of $^3$He in the event. However, no type III radio bursts (Fig.~\ref{f8}, Mid), a signature of energetic electron production and escape, nor energetic electrons (see Fig.~\ref{f8}, Bottom) were observed in association with the jet. Remember, the type III radio burst and the electron event observed by Solar Orbiter are significantly later than the start of the jet. It is known that $^3$He-rich SEP events show a high ($\sim$99\%) association with type III radio bursts \citep[e.g.,][]{2006ApJ...650..438N}, and only a few cases have been reported without type III bursts \citep{2009ApJ...700L..56M}. Thus, the production of $^3$He in the jet-like eruption is unlikely.

On its way from the Sun, the event-associated CME shock wave can accelerate remnant $^3$He from a heliospheric reservoir. The reservoir creation requires $^3$He injections and preceding magnetic disturbances, such as interplanetary CMEs, that inhibit particle escape \citep[e.g.,][]{1992GeoRL..19.1243R,2013SSRv..175...53R}. The long period of energetic $^3$He on 2020 November 17--23 (Fig.~\ref{f3} left)  was caused by at least five ion injections; three on November 17, one on November 18, and one on November 20 \citep{2021A&A...656L..11B}. The responsible source rotated from E90 to E50 (viewed from STEREO-A), thus filling a substantial volume in the heliosphere with $^3$He. The median values of $^3$He/$^4$He and Fe/O at 0.2--2.0\,MeV\,nucleon$^{-1}$ from these injections are 0.56 and 0.91, respectively. The $^3$He intensity gradually declines before the end of November 23. Using the LASCO CME catalog, we searched for November’s CMEs that preceded these ion injections. We looked for CMEs with angular widths $\geq$60\degr and the position angle (PA; measured from north counter-clockwise) of 0\degr--33\degr or 213\degr--360\degr. The CMEs with such PAs would travel toward the hemisphere where Solar Orbiter is located. We find six CMEs with such criteria. These CMEs, whose linear speed varies between $\sim$100 and $\sim$300\,km\,s$^{-1}$, would travel to distances of 0.3--1.8\,au at the liftoff time of the CME associated with the SEP event studied here. Therefore, the medium was appropriate for the confinement of $^3$He ions from prior events. However, the CME source AR 12790 lies about 100$\degr$  from the AR 12784/786 complex that produced the impulsive $^3$He-rich events.  Since the CME made the SEP event on PSP, whose magnetic foot-point is located further west from the AR12786/785 complex (see Fig.~\ref{f7}), we suppose a broad shock from the CME extending well to the west into the reservoir accelerated the $^3$He seed particles.

In order to determine whether $^3$He was produced in the parent AR or other sites on the Sun we indicate that the examined event was accompanied by a type III radio burst and the causative electrons, commonly seen in $^3$He-rich SEP events. Unfortunately, we lack EUV imaging observation for a detailed examination of a possible jet activity behind the limb (from the STEREO-A perspective). \citet{2006JGRA..111.6S90C,2010JGRA..115.8101C} pointed out that the presence of type III radio emission in GSEP events implies that ions accelerated in a concomitant flare can escape into interplanetary space. A direct magnetic connection of Solar Orbiter to the parent AR (see Fig.~\ref{f7}) could allow ion escape and their measurement in the November 24 event. Previously, \citet{1991ApJ...373..675C} suggested that enhanced heavy-ion abundances measured in well-connected GSEP events had a component from flare-accelerated particles. In the examined event, the Fe/O is consistent with GSEP events. The obvious temporal correlation between the $^3$He and other elements shown in detail in Fig.~\ref{f4} indicates that the increased intensities are related to a single source. To argue that the $^3$He somehow appeared with just this timing and intensity as a result of independent activities requires constructing a very unlikely scenario. Therefore, we conclude the most likely origin of the observed $^3$He enhancement is due to remnant material from a preceding long period of $^3$He-rich SEP injections. 

A noteworthy feature in the examined period was the storm type III bursts observed by STEREO-A. It has been suggested that storm type III bursts \citep{1970SoPh...15..222F} are caused by trapped solar energetic electron beams on the coronal loops \citep[e.g.,][]{2004ASSL..314..305G}. It seems improbable that these closed sites provide suprathermal $^3$He for accelerations in GSEP events.

\begin{acknowledgements}
R.~Bu\v{c}\'ik was supported by NASA grant No. 80NSSC21K1316 and NASA contract NNN06AA01C. V.~Krupar acknowledges the support by NASA under grants 18-2HSWO2182-0010 and 19-HSR-192-0143. Solar Orbiter is a mission of international cooperation between ESA and NASA, operated by ESA. The Suprathermal Ion Spectrograph (SIS) is a European facility instrument funded by ESA. The SIS instrument was constructed by the Johns Hopkins Applied Physics Lab. and CAU Kiel.  Post launch operation of SIS at APL is funded by NASA contract NNN06AA01C. The UAH team acknowledges the financial support of the Spanish MINCIN Project PID2019-104863RBI00/AEI/10.13039/501100011033. We thank the German Space Agency, DLR, for the build and support of STEP, EPT, and HET with grants 50OT0901, 50OT1202, 50OT1702, and 50OT2002. Solar Orbiter magnetometer data was provided by Imperial College London and supported by the UK Space Agency. The LASCO C2 CME catalog is generated and maintained at the CDAW Data Center by NASA and The Catholic University of America in cooperation with the Naval Research Laboratory. SOHO is a project of international cooperation between ESA and NASA.
\end{acknowledgements}

%
   \bibliographystyle{aa} 
   \bibliography{ads} 
%

\end{document}